\documentclass[
    twocolumn,
	prd,
	amssymb,
	preprintnumbers,superscriptaddress,
	nofootinbib]{revtex4-1}

\pdfoutput=1
\usepackage{paralist}
\usepackage{graphicx}
\usepackage{enumitem}
\usepackage{latexsym}
\usepackage{amsfonts}
\usepackage{amssymb}
\usepackage{xcolor}
\usepackage[export]{adjustbox}
\usepackage{amsmath}
\usepackage[thinlines]{easytable}
\usepackage{slashed}
\usepackage{dcolumn}
\usepackage{verbatim}
\usepackage{float}
\usepackage{multirow}
\usepackage{xspace}
\usepackage[normalem]{ulem}
\usepackage[
pdfauthor={Jeremy Sakstein}]{hyperref}
\usepackage{tabularx}
\usepackage{lettrine}
\usepackage{setspace}

\input Zallman.fd

\LettrineTextFont{\itshape}

\renewcommand{\tilde}{\widetilde} 

\newcommand{\beq}{\begin{equation}}
\newcommand{\eeq}{\end{equation}}
\newcommand{\bea}{\begin{eqnarray}}
\newcommand{\eea}{\end{eqnarray}}

\usepackage{pdfbase}[2017/03/16]
\usepackage{xparse,ocgbase}
\usepackage{xcolor,calc}
\usepackage{tikzpagenodes,linegoal}
\usetikzlibrary{calc}
\usepackage{tcolorbox}

\ExplSyntaxOn
\let\tpPdfLink\pbs_pdflink:nn
\let\tpPdfAnnot\pbs_pdfannot:nnnn\let\tpPdfLastAnn\pbs_pdflastann:
\let\tpAppendToFields\pbs_appendtofields:n
\def\tpPdfXform{\pbs_pdfxform:nnnnn{1}{1}{}{}}
\let\tpPdfLastXform\pbs_pdflastxform:
\let\cListSet\clist_set:Nn\let\cListItem\clist_item:Nn
\ExplSyntaxOff

\usepackage{pdfbase}[2017/03/16]
\usepackage{xparse,ocgbase}
\usepackage{xcolor,calc}
\usepackage{tikzpagenodes,linegoal}
\usetikzlibrary{calc}
\usepackage{tcolorbox}

\ExplSyntaxOn
\let\tpPdfLink\pbs_pdflink:nn
\let\tpPdfAnnot\pbs_pdfannot:nnnn\let\tpPdfLastAnn\pbs_pdflastann:
\let\tpAppendToFields\pbs_appendtofields:n
\def\tpPdfXform{\pbs_pdfxform:nnnnn{1}{1}{}{}}
\let\tpPdfLastXform\pbs_pdflastxform:
\let\cListSet\clist_set:Nn\let\cListItem\clist_item:Nn
\ExplSyntaxOff

\makeatletter
\NewDocumentCommand{\tooltip}{%
  ssssO{\ifdefined\@linkcolor\@linkcolor\else blue\fi}mO{yellow!20}mO{0pt,0pt}%
}{{%
  \leavevmode%
  \IfBooleanT{#2}{%
    \ocgbase@new@ocg{tipOCG.\thetcnt}{%
      /Print<</PrintState/OFF>>/Export<</ExportState/OFF>>%
    }{false}%
    \xdef\tpTipOcg{\ocgbase@last@ocg}%
    \ocgbase@add@ocg@to@radiobtn@grp{tool@tips}{\ocgbase@last@ocg}%
  }%
  \tpPdfLink{%
    \IfBooleanTF{#4}{%
      /Subtype/Link/Border[0 0 0]/A <</S/SetOCGState/State [/Toggle \tpTipOcg]>>
    }{%
      /Subtype/Screen%
      /AA<<%
        \IfBooleanTF{#3}{%
          /E<</S/SetOCGState/State [/Toggle \tpTipOcg]>>%
        }{%
          \IfBooleanTF{#2}{%
            /E<</S/SetOCGState/State [/ON \tpTipOcg]>>%
            /X<</S/SetOCGState/State [/OFF \tpTipOcg]>>%
          }{
            \IfBooleanTF{#1}{%
              /E<</S/JavaScript/JS(%
                var fd=this.getField('tip.\thetcnt');%
                if(typeof(click\thetcnt)=='undefined'){%
                  var click\thetcnt=false;%
                  var fdor\thetcnt=fd.rect;var dragging\thetcnt=false;%
                }%
                if(fd.display==display.hidden){%
                  fd.delay=true;fd.display=display.visible;fd.delay=false;%
                }else{%
                  if(!click\thetcnt&&!dragging\thetcnt){fd.display=display.hidden;}%
                  if(!dragging\thetcnt){click\thetcnt=false;}%
                }%
                this.dirty=false;%
              )>>%
            }{%
              /E<</S/JavaScript/JS(%
                var fd=this.getField('tip.\thetcnt');%
                if(typeof(click\thetcnt)=='undefined'){%
                  var click\thetcnt=false;%
                  var fdor\thetcnt=fd.rect;var dragging\thetcnt=false;%
                }%
                if(fd.display==display.hidden){%
                  fd.delay=true;fd.display=display.visible;fd.delay=false;%
                }%
               this.dirty=false;%
              )>>%
              /X<</S/JavaScript/JS(%
                if(!click\thetcnt&&!dragging\thetcnt){fd.display=display.hidden;}%
                if(!dragging\thetcnt){click\thetcnt=false;}%
                this.dirty=false;%
              )>>%
            }%
            /U<</S/JavaScript/JS(click\thetcnt=true;this.dirty=false;)>>%
            /PC<</S/JavaScript/JS (%
              var fd=this.getField('tip.\thetcnt');%
              try{fd.rect=fdor\thetcnt;}catch(e){}%
              fd.display=display.hidden;this.dirty=false;%
            )>>%
            /PO<</S/JavaScript/JS(this.dirty=false;)>>%
          }%
        }%
      >>%
    }%
  }{{\color{#5}#6}}%
  \sbox\tiptext{%
    \IfBooleanT{#2}{%
      \ocgbase@oc@bdc{\tpTipOcg}\ocgbase@open@stack@push{\tpTipOcg}}%
    \tcbox[colframe=black,colback=#7,size=fbox,arc=1ex,sharp corners=southwest]{#8}%
    \IfBooleanT{#2}{\ocgbase@oc@emc\ocgbase@open@stack@pop\tpNull}%
  }%
  \cListSet\tpOffsets{#9}%
  \edef\twd{\the\wd\tiptext}%
  \edef\tht{\the\ht\tiptext}%
  \edef\tdp{\the\dp\tiptext}%
  \tipshift=0pt%
  \IfBooleanTF{#2}{%
    \setlength\whatsleft{\linegoal}%
  }{%
    \measureremainder{\whatsleft}%
  }%
  \ifdim\whatsleft<\dimexpr\twd+\cListItem\tpOffsets{1}\relax%
    \setlength\tipshift{\whatsleft-\twd-\cListItem\tpOffsets{1}}\fi%
  \IfBooleanF{#2}{\tpPdfXform{\tiptext}}%
  \raisebox{\heightof{#6}+\tdp+\cListItem\tpOffsets{2}}[0pt][0pt]{%
    \makebox[0pt][l]{\hspace{\dimexpr\tipshift+\cListItem\tpOffsets{1}\relax}%
    \IfBooleanTF{#2}{\usebox{\tiptext}}{%
      \tpPdfAnnot{\twd}{\tht}{\tdp}{%
        /Subtype/Widget/FT/Btn/T (tip.\thetcnt)%
        /AP<</N \tpPdfLastXform>>%
        /MK<</TP 1/I \tpPdfLastXform/IF<</S/A/FB true/A [0.0 0.0]>>>>%
        /Ff 65536/F 3%
        /AA <<%
          /U <<%
            /S/JavaScript/JS(%
              var fd=event.target;%
              var mX=this.mouseX;var mY=this.mouseY;%
              var drag=function(){%
                var nX=this.mouseX;var nY=this.mouseY;%
                var dX=nX-mX;var dY=nY-mY;%
                var fdr=fd.rect;%
                fdr[0]+=dX;fdr[1]+=dY;fdr[2]+=dX;fdr[3]+=dY;%
                fd.rect=fdr;mX=nX;mY=nY;%
              };%
              if(!dragging\thetcnt){%
                dragging\thetcnt=true;Int=app.setInterval("drag()",1);%
              }%
              else{app.clearInterval(Int);dragging\thetcnt=false;}%
              this.dirty=false;%
            )%
          >>%
        >>%
      }%
      \tpAppendToFields{\tpPdfLastAnn}%
    }%
  }}%
  \stepcounter{tcnt}%
}}
\makeatother
\newsavebox\tiptext\newcounter{tcnt}
\newlength{\whatsleft}\newlength{\tipshift}
\newcommand{\measureremainder}[1]{%
  \begin{tikzpicture}[overlay,remember picture]
    \path let \p0 = (0,0), \p1 = (current page.east) in
      [/utils/exec={\pgfmathsetlength#1{\x1-\x0}\global#1=#1}];
  \end{tikzpicture}%
}

\newcommand{\msun}{{\rm M}_\odot}
 
\newcommand{\mdm}{m_{\rm DM}}

\newcommand{\dcedit}[1]{\textcolor{black}{#1}}

\DeclareRobustCommand{\okina}{%
  \raisebox{\dimexpr\fontcharht\font`A-\height}{%
    \scalebox{0.8}{`}%
  }%
}

\allowdisplaybreaks

\begin{document}

\title{Dark Matter Annihilation and Pair-Instability Supernovae}

\author{Djuna Croon} \email{djuna.l.croon@durham.ac.uk}
\affiliation{Institute for Particle Physics Phenomenology, Department of Physics, Durham University, Durham DH1 3LE, U.K.}
\author{Jeremy Sakstein} \email{sakstein@hawaii.edu}
\affiliation{Department of Physics \& Astronomy, University of Hawai\okina i, Watanabe Hall, 2505 Correa Road, Honolulu, HI, 96822, USA}

\date{\today}

\begin{abstract}
We study the evolution of heavy stars ($M\ge40\msun$) undergoing pair-instability in the presence of annihilating dark matter.~Focusing on the scenario where the dark matter is in capture-annihilation equilibrium, we model the profile of energy injections in the local thermal equilibrium approximation.~We find that significant changes to masses of astrophysical black holes formed by (pulsational) pair-instability supernovae can occur when the ambient dark matter density $ \rho_{\rm DM} \gtrsim10^9 \rm \, GeV \, cm^{-3}$.~There are two distinct outcomes, depending on the dark matter mass.~For masses $m_{\rm DM}\gtrsim1$ GeV the DM is primarily confined to the core.~The annihilation increases the lifetime of core helium burning, resulting in more oxygen being formed, fueling a more violent explosion during the pair-instability-induced contraction.~This drives stronger pulsations, leading to lighter black holes being formed than predicted by the standard model.~For masses $m_{\rm DM}\lesssim0.5$ GeV there is significant dark matter in the envelope, leading to a phase where the star is supported by the energy from the annihilation.~This reduces the core temperature and density, allowing the star to evade the pair-instability allowing heavier black holes to be formed.~We find a mass gap for all models studied.
\end{abstract}

\preprint{IPPP/23/62}

\maketitle

\section{Introduction}

The advent of gravitational wave astronomy has revolutionised the study of compact objects.~With the publication of catalogues of transient (merger) events, the population statistics of astrophysical black holes can be studied for the first time.~This information enables us to probe the processes governing the massive star progenitors of these black holes, allowing us to test stellar structure theory.~For example, it is possible to measure the rates of nuclear reactions, most importantly the $^{12}{\rm C}(\alpha,\gamma)^{16}{\rm O}$ reaction \cite{Farmer:2019jed,Farmer:2020xne,Mehta:2021fgz}.~In addition, new physics beyond the Standard Model (BSM) can alter the structure, evolution, and formation of these objects either via their effects on the evolution of the stars through the pulsational pair-instability supernovae (PPISN) and pair-instability supernovae (PISN) phases \cite{Croon:2020ehi,Croon:2020oga,Straight:2020zke,Sakstein:2020axg,Ziegler:2020klg,Sakstein:2022tby,Fernandez:2022zmc,Ziegler:2022apq}, or via their effects on the formation of black hole binaries \cite{Croon:2022rrv}.~Examples of the new physics studied include light ($\mdm\lesssim 10$ keV) weakly interacting particles, intermediate mass particles ($\mdm\lesssim 1$ GeV) strongly coupled to the Standard Model (SM), and dark energy.~Continuing this program, in this work we study the effects of dark matter (DM) with $\mdm\sim \mathcal{O}(1)$ GeV.~In this mass range, the DM is too heavy to be thermally produced by the stellar plasma, but can be gravitationally captured if the ambient dark matter density is sufficiently large.~The subsequent annihilation results in an injection of energy into the stellar material.~

The results of our study can bu summarized as follows:
\begin{enumerate}
    \item When DM is heavier than $\sim1$ GeV it is highly concentrated in the star's core.~The energy injection from annihilation augments the lifetime of core helium burning because less energy is required to maintain hydrostatic equilibrium.~This gives more time for the $^{12}{\rm C}(\alpha,\gamma)^{16}{\rm O}$ reaction to operate, resulting in a larger $^{16}$O abundance at core helium depletion.~This drives a more violent explosion, which causes the star to lose more mass.
    ~This results in a reduction in the location of the lower edge of the upper black hole mass gap.
    \item When DM is lighter than $\sim1$  GeV it is more diffuse, injecting energy thought the entire star.~Energy  injected into non-nunclear burning regions post-helium depletion drives a  phase of evolution where the star is partly supported by DM annihilation.~During this phase, the star's central density and temperature decrease, allowing it to evade the pair-instability in the $\rho$--$T$ plane.~Stars that would have otherwise undergone PPISN instead directly core collapse, forming heavier black holes than predicted in the absence of DM annihilation.~This effect vanishes in heavier stars which require more energy for support than is be provided by DM annihilation, and these objects experience PPISN and PISN, resulting in a mass gap.
\end{enumerate}
We exemplify the scenarios above by studying two DM masses: $\mdm=1$ GeV and $\mdm=0.2$ GeV.~Our results differ from those obtained in earlier work on dark heat injections and PPISN \cite{Ziegler:2020klg,Ziegler:2022apq}.~We discuss this discrepancy below.

\begin{spacing}{0.96}
This paper is organized as follows.~In section \ref{sec:DM_injection} we describe the DM model we adopt.~In section \ref{sec:stellar_response} we describe our implementation of DM annihilation into our stellar structure code.~Our results are presented in section~\ref{sec:results}.~We discuss our results and conclude in section \ref{sec:conclusions}.
\end{spacing}

\section{Dark matter energy injections}
\label{sec:DM_injection}
To model the energy injections due to DM annihilation, we must first consider the appropriate dark matter number density profile in the star.~This is a complicated procedure that requires a careful consideration of the effects of DM capture, annihilation, and evaporation --- all of which depend on the size and structure of the object under consideration, the timescale for this object's evolution, and the DM mass.~Since our study is the first of its kind, we will make a number of simplifying assumptions that enable us to investigate the effects of DM annihilation across a wide range of masses and over the entire evolution of the star.~One goal of this work is to determine whether a more comprehensive study is required, and over which range of DM masses.~These approximations facilitate this.

First, we assume the dark matter is thermalised in the core.~The thermalisation time can be estimated using the number of scatterings needed (given by the DM-SM mass ratio) times the DM mean free path ($\lambda =(\sigma_0 n_{\rm SM})^{-1} $, where $\sigma_0$ is the cross section for DM-SM scattering) divided by the DM dispersion velocity in the star \cite{Iocco:2008rb}.~Taking the latter to be the escape velocity at the surface (or the surface of the core), one obtains
\begin{equation}
\begin{split}
    \tau_{\rm th}(r) =& \frac{m_{\rm DM}}{\rho_{\rm SM}(r) \sigma_0} \sqrt{\frac{r}{2 G M}} \\
        =& 1.9 \, {\rm s} \left(\frac{m_{\rm DM}}{1 \rm GeV} \right)
        \left( \frac{\sigma_0}{10^{-35} \rm cm^{2}}\right)^{-1} \\&\times
        \left( \frac{R_\star}{0.5 R_\odot}\right)^{7/2}
        \left(\frac{M_\star}{10 M_\odot} \right)^{-3/2}.
    \end{split}
    \label{eq:tauth}
\end{equation}
where the last equality approximates the star as having uniform density.~We note that the latter approximation is for illustrative purposes only. We expect that the result --- that thermalisation happens on much smaller timescales than the typical timescales of stellar evolution --- continues to hold true for more realistic density profiles.

Our second approximation is that annihilation equilibrium has been established on timescales larger than $\tau_{\rm eq} \sim \left( \Gamma_{\rm ann} \Gamma_{\rm cap} \right)^{-1/2} $.
The capture rate we can estimate as $\Gamma_{\rm cap} = \Phi \pi R^2$ where $\Phi $ is the flux density of DM which is captured, given by \cite{Leane:2022hkk}
\begin{equation}
\label{eq:DMflux}
    \Phi = v_{\rm DM} \sqrt{\frac{8}{3\pi}} \left[ 1 + \frac{3}{2}\left(\frac{v_{\rm esc}}{v_{\rm DM}}\right)^2 \right] \frac{\rho_{\rm DM} f_{\rm cap}}{m_{\rm DM}} 
\end{equation}
where $f_{\rm cap}$ denotes the fraction of DM which is captured as it streams through the star, and where $\rho_{\rm DM}$ is the DM halo energy density.~We can estimate the DM annihilation rate per effective volume as
$\Gamma_{\rm ann} =  \langle \sigma_{\rm ann} v \rangle / V_{\rm eff}$, with $1/V_{\rm eff} = \int_V n_\chi^2/\int_V n_\chi$ as in \cite{Croon:2023bmu}.
This gives 
\begin{equation}
\begin{split}
    \tau_{\rm eq} \sim& \left( \Gamma_{\rm ann} \Gamma_{\rm cap} \right)^{-1/2} 
    \\ 
    \sim& 3 \times 10^4 {\rm yr} \sqrt{
    \frac{10^{-5}}{f_{\rm cap}}
    \frac{ m_\chi \, 10^{-30} \text{ cm}^3/\text{s} }{ \langle \sigma_{\rm ann} v\rangle \text{ GeV} }}
\end{split}
\label{eq:taueq}
\end{equation}
where we have assumed an ambient DM density of $10^9 \, \rm GeV \, cm^{-3}$, motivated by the results we present below that show deviations from the SM emerging for these values.~Using {\tt Asteria} \cite{Leane:2023woh}, we find that the reference capture fraction $f_{\rm cap} = 10^{-5}$ is relevant for DM-nucleon cross sections of $\sigma_0 \lesssim 10^{-42}\, \rm cm^{2} $.~For $\sigma_0 = 10^{-35}\, \rm cm^{2} $, $f_{\rm cap}$ approaches unity.~To maximize the potential signal, we take $f_{\rm cap}=1$ below, implying that our results hold for $\sigma_0 \gtrsim 10^{-35}\, \rm cm^{2} $.

Our third approximation, implicit in the approximation that the DM is in capture-annihilation equilibrium, is that evaporation is negligible.~This is approximately valid for DM masses for which the evaporation rate is below the capture rate.~We estimate the evaporation rate as 
$ \Gamma_{\rm evap} = 4 \pi \Phi_{\rm J} (T) R_{\star}^2$, where $\Phi_{\rm J} (T)$ is the Jeans flux 
\begin{equation}
\label{eq:jeansflux}
    \Phi_{\rm J} (T) =  \frac{n v}{2 \sqrt{\pi}} \left(1 + \frac{v_{\rm esc}^2}{v^2} \right) \exp \left(- \frac{v_{\rm esc}^2}{v^2} \right)
\end{equation}
with $n$  the DM number density in the star, $v=\sqrt{2 T/m_{\rm DM}}$, and $v_{\rm esc} = \sqrt{2 G M/R }$ being the surface escape velocity.~This leads to evaporation masses on the order of $m_{\rm evap} \sim \mathcal{O } (0.1)$ GeV for $ 65 \msun$ stars, with a mild dependence on the cross section in the local thermal equilibrium regime (see below).~Note that the evaporation mass can be lower in the presence of long-range forces in the dark sector \cite{Acevedo:2023owd}. \dcedit{The inclusion of such forces would lead to a slight modification of the profiles below.}

Our fourth assumption is that the DM diffusion through the star is fast compared to both the timescales on which the total DM particle number significantly changes, and the stellar evolution.~The diffusive timescale can be estimated as $t_{\rm diff} \sim (\Delta x)^2 n_{\rm SM} \sigma_0/ v_T 
\sim \rm \mathcal{O}(few) \, min $ ~\cite{Leane:2022hkk}.~The advective timescale for these stars is typically much larger.~Turning to the stellar evolution timescales, the relevant timescales in our simulation are small compared to \eqref{eq:tauth} and \eqref{eq:taueq} with the exception of the pulsations which can occur on timescales $ \mathcal{O}(\rm days)$.~If annihilation equilibrium would be assumed at this stage the energy injection would grow with the increase in radius of the star, resulting in a PISN whenever a PPISN is triggered.~To conservatively model the behaviour of DM for smaller cross sections, we freeze its profile at the onset of the first pulsation.~

The spatial profile that the DM assumes in the star depends on the \textit{Knudsen number}.~In the limit of small Knudsen numbers --- that is, dark matter scatterings are frequent;~the characteristic mean free path $\lambda$ is short compared to the core radius $\sim 0.5 R_\odot $ as well as compared to the scale of temperature variations $ |\nabla\ln T|^{-1}$ --- the Local Thermal Equilibrium (LTE) assumption holds.~We find that a $72 \msun$ star has $ |\nabla\ln T|^{-1} > \mathcal{O}(10^{-1}) \rm R_\odot$ in the core on the ZAHB and at the onset of the first pulsation.~This means that $ \sigma_0 n_{\rm SM} \geq 10^{-11} \rm cm^{-1}$ defines the short mean free path regime.~For a number density of $ 10^{25} \, \rm cm^{-3}$, this corresponds to $ \sigma_0 =10^{-36} \, \rm cm^2$.
In the LTE regime, the DM profile $n_{\rm DM}$ in the star is given by the Maxwell-Boltzmann distribution \cite{Gould:1989hm}: 
\begin{equation}\label{eq:MBdist}
    \left(\frac{n_{\rm DM} (r) }{n_{\rm DM}(0)} \right)=
    \left(\frac{T(r)}{T(0)} \right)^{3/2} 
    e^{- \int_0^r d\tilde{r} 
    \frac{\alpha(\tilde{r})dT/d\tilde(r) + m_{\rm DM} g(\tilde{r}) }{T(\tilde{r})}
    }
\end{equation}
Here 
$g(r)$ is the gravitational acceleration and
$\alpha (r)$  is a ``separation constant", related to the diffusion coefficient in \cite{Leane:2022hkk} by $\kappa = \alpha - 5/2$.~Equation \eqref{eq:MBdist} can be simplified in the limit that $\alpha$ is independent of $r$ --- in this limit there also is a simple analytic expression which matches the numerical result closely \cite{Leane:2022hkk}, such that 
\begin{equation}\label{eq:MBdist2}
    \left(\frac{n_{\rm DM} (r) }{n_{\rm DM}(0)} \right)= 
    \left(\frac{T(r)}{T(0)} \right)^{-1 + \frac{1}{2} \left(1+\frac{m_{\rm DM}}{m_{\rm SM}}\right)^{-3/2}} 
    e^{- \int_0^r d\tilde{r} 
    \frac{ m_{\rm DM} g(\tilde{r}) }{T(\tilde{r})}
    }.
\end{equation}
In the opposite limit of large Knudsen number, the profile is expected to take on the isothermal distribution derived by \cite{Spergel:1984re}.~We leave an analysis of that regime for future work.~The energy injection (per unit mass and time) is then given by 
\begin{equation}
\label{eq:energyinjection}
\begin{split}
    \epsilon_{\rm DM} = f_\nu \frac{\langle \sigma v \rangle n_{\rm DM}^2(r) m_{\rm DM}}{\rho(r)}
\end{split}
\end{equation}
where
we substitute in the profile \eqref{eq:MBdist2}, $\rho(r)$ is the density of the star at radius $r$, $ \langle \sigma v \rangle$ is the annihilation cross section, and
the dimensionless factor $f_\nu$ accounts for annihilation biproducts that free-stream out of the star.~As we are interested in the maximal effect, we will take $f_\nu = 1$.~The DM annihilation luminosity in annihilation equilibrium is given by
\begin{equation}
    L_{\rm DM} = 4 \pi \int_0^{R_\star} \rho(r) \epsilon_{\rm DM} (r) r^2 dr = m_{\rm DM} \Gamma_{\rm cap}
    \label{eq:lumanneq}
\end{equation}
from which it is seen that the energy injection must be independent of $\langle \sigma v \rangle$ in annihilation equilibrium \cite{Lopes:2021jcy}.
We can now normalise our energy injection profile \eqref{eq:energyinjection} using \eqref{eq:lumanneq}.

\begin{figure}
    \centering
    \includegraphics[width=.45\textwidth]{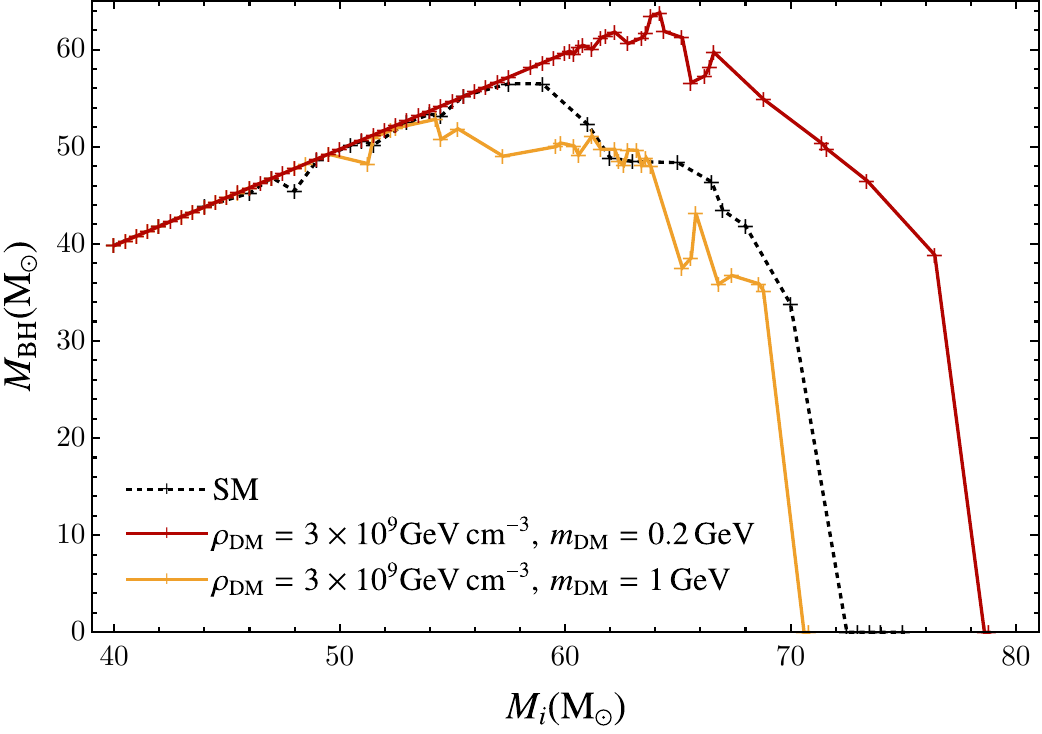}
    \caption{Results of DM injections described in text for two different DM masses.}
    \label{Fig:gridLTE}
\end{figure}

\section{Stellar Structure Code}
\label{sec:stellar_response}
We modified the stellar structure code MESA version 12778 \cite{Paxton:2010ji,Paxton:2013pj,Paxton:2015jva,Paxton:2017eie} to include dark matter injection with the profile \eqref{eq:MBdist2}.~This was accomplished by using the {\tt other\_energy\_implict} hook.~The full details of our implementation can be found in a reproduction package that accompanies this paper:~\href{https://zenodo.org/records/10056048}{https://zenodo.org/records/10056048}.~This  includes MESA inlists and modifications.~We briefly summarize the salient features of our stellar modeling choices here for completeness, referring the reader to \cite{sakstein_2023_10056048} for the full details.~Our simulations include convection described by mixing length theory, semiconvection, and overshooting;~all with parameters given in the inlists in the reproduction package \cite{sakstein_2023_10056048}.~We use the mass-loss scheme of \cite{Brott:2011ni}, which is a good description of wind-loss for massive stars.~Our treatment of the pulses follows the prescription of references \cite{Farmer:2019jed,Croon:2020oga}, to which we refer the reader for the details.~All nuclear reaction rates are set to the MESA defaults described in \cite{Farmer:2019jed,Croon:2020oga} with the exception of the $^{12}{\rm C}(\alpha,\gamma)^{16}{\rm O}$ rate.~This is known to be the largest source of uncertainty in the location of the black hole mass gap and the shape of the black hole mass spectrum, so we use the state-of-the-art rate table presented in \cite{Mehta:2021fgz}\footnote{We are grateful to J.~de Boer for providing this table.}.~Our reproduction package contains a copy of this table \cite{sakstein_2023_10056048}.

Our implementation of the DM annihilation is as follows.~We write equation \eqref{eq:energyinjection} (having substituted equation \eqref{eq:MBdist2}) as
\begin{equation}\label{eq:eps(r)redef}
    \varepsilon(r)=\varepsilon_0\frac{f(r)^2}{\rho(r)}
\end{equation}
with
\begin{align}
    \varepsilon_0&=\langle\sigma v\rangle \mdm n^2_{\rm DM}(0);\quad \textrm{and}\\
    f(r)&=\left(\frac{T(r)}{T(0)} \right)^{-1 + \frac{1}{2} \left(1+\frac{m_{\rm DM}}{m_{\rm SM}}\right)^{-3/2}} 
    e^{- \int_0^r d\tilde{r} 
    \frac{ m_{\rm DM} g(\tilde{r}) }{T(\tilde{r})}
    }.
\end{align}
With these definitions, equation \eqref{eq:lumanneq} can be rearranged to find $\varepsilon_0$
\begin{equation}\label{eq:epsilon0def}
    \varepsilon_0=\frac{R^2\Phi\mdm}{4\chi};\quad\chi=\int_0^{R_*}r^2f(r)^2\mathrm{d}r,
\end{equation}
where we took $f_\nu=1$, $\Gamma_{\rm cap}=\pi R^2\Phi$, and $\Phi$ is given by equation \eqref{eq:DMflux}.~At each time step, we calculate $\chi$ by integrating over the stellar profile and use equation \eqref{eq:epsilon0def} to calculate $\varepsilon_0$.~We then use this to calculate the injection in each cell using \eqref{eq:eps(r)redef}.~

 \section{Results}
 \label{sec:results}

\subsection{Theoretical Considerations}

Before proceeding to the numerical results, one can gain some insight into the effects of DM annihilation from purely theoretical considerations.~The initial question of  interest is:~can the injection alter the location of the tracks in the $\rho$--$T$ plane?~The answer reveals whether the star will encounter the pair-instability.~The shape of the stellar tracks in the $\rho$--$T$ plane can be found by assuming that the star is radiation-supported with equation of state (EOS) $P\propto T^4$, which is reasonable for massive objects.~In this case, the equation for hydrostatic equilibrium gives \cite{Straight:2020zke}
\begin{equation}
\log(\rho)=\frac13\log(T)+\frac16\log(M)+c,
\end{equation}
where $c$ is a constant that depends on Newton's constant and the radiation constant.~A similar relation but with a different slope can be derived for the case of gas pressure domination where $P\propto\rho T$, with the true EOS being a combination of the two.~Thus we see that the slope of the tracks depend only on the equation of state, and therefore the DM injection only alters the track if it is sufficiently strong  that the EOS is significantly modified.~In terms of the DM mass, the profile \eqref{eq:MBdist2} implies that for light DM the injection is diffuse throughout the star and can inject energy into regions where there is no nuclear burning and therefore dominate the EOS.~For heavier DM, the injection is concentrated in the core, and we observe that the stellar tracks are unmodified before the onset of pair instability.

\subsection{Numerical Results}

We simulated a grid of stars with masses in the range $40\msun\le M\le 90 \msun$ from the onset of helium burning through the regime of pair-instability to their ultimate core collapse or PISN, depending on the mass.~We took the metallicity to be $Z=10^{-5}$ corresponding to population-III stars.~This choice was made because low metallicity objects lose less mass to stellar winds compared with their high metallicity counterparts and, consequentially, form heavier BHs  \cite{Farmer:2019jed}.~These objects therefore set the location of the lower edge of the upper mass gap, which is the observable targeted by this study.~Two different DM masses were investigated, $\mdm=1$ GeV and $\mdm=0.2$ GeV for a range of ambient DM densities $\rho_{\rm DM}$.~We fixed the DM velocity to be $v_{\rm DM}=220$ km/s corresponding to the circular velocity of the Solar System through the Milky Way.~Our reproduction package includes the option of varying this \cite{sakstein_2023_10056048}, but we chose not to do so since this is nearly degenerate with $\rho_{\rm DM}$.~The degeneracy is only lifted by the $\mathcal{O}(1)$ \textit{gravitational focusing} correction to the flux in equation \eqref{eq:DMflux}.~In addition, the DM density is expected to vary over many orders-of-magnitude, whereas we expect only an  $\mathcal{O}(1)$ variation of $50\textrm{km/s}\lesssim v_{\rm DM}\lesssim 300\textrm{km/s}$ given the distribution of circular velocities of galaxies.~We therefore expect that the ambient DM density is the most important factor  determining the onset of the effects of DM annihilation in stars.

We varied $0.42\,\textrm{GeV cm}^{-3}\le\rho_{\rm DM}\le 5\times10^{9}\,\textrm{GeV cm}^{-3}$ i.e., spanning DM densities from that of the solar neighborhood to the largest densities one would expect for stars near the galactic center in scenarios with a DM spike \cite{Gondolo:1999ef} (see \cite{Shen:2023kkm} for recent observational constraints on such a spike in the Milky Way based on the trajectories of the nearest stars).~Such a large exploration was necessary because the DM energy injection is a complicated function of the stellar structure, temperature, and density, and, similarly, the energy released from nuclear burning is a strong function of temperature and density, making a theoretical comparison difficult.~We found deviations from the SM for $\rho_{\rm DM}\gtrsim10^9$  GeV/cm$^3$.~Our results are shown in figure \ref{Fig:gridLTE}, where we plot the final black hole mass as a function of zero-age helium burning (ZAHB) mass  for the SM and both DM cases we investigated.~The two masses studied show opposite effects.~We elucidate each of these in turn below.

\begin{figure}
    \centering
    \includegraphics[width=.45\textwidth]{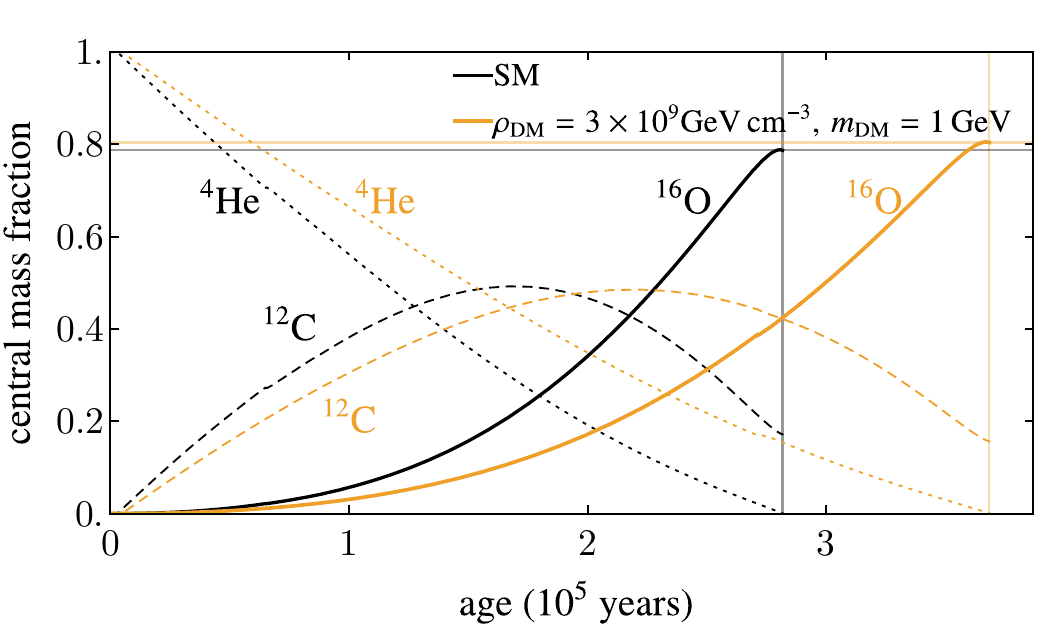}
    \caption{
    Core elemental abundances for a $59 \msun$ star.~The $^{16}$O abundance at helium depletion is indicated by the horizontal lines.~As explained in text, the DM heat injection slows down the evolution and leads to a higher core oxygen abundance, exacerbating pair instability.
    }
    \label{Fig:CtoO}
\end{figure}

The inclusion of annihilating DM with mass $\mdm=1$ GeV results in stronger PPISN, and a reduction in the ZAHB mass at which the PPISN/PISN transition occurs.~This leads to lower black hole masses, and a reduction in the location of the lower edge of the upper black hole mass gap.~DM with this mass is concentrated in the core.~The energy released from its annihilation prolongs the lifetime of core helium burning because less nuclear fuel needs to be burned to stave off gravitational collapse.~During this time, the $^{12}{\rm C}(\alpha,\gamma)^{16}{\rm O}$ reaction reprocesses some of the carbon into oxygen.~The increased lifetime provides more time for this process, ultimately enhancing the amount of oxygen produced at core helium depletion.~Both of these effects are exemplified in Fig.~\ref{Fig:CtoO} for a $59\msun$ star.~The increase in core oxygen levels drives in a more violent explosion.~This results in enhanced mass loss for stars that undergo PPISN, as seen in Fig.~\ref{Fig:gridLTE}.

\begin{figure}
    \centering
    \includegraphics[width=.45\textwidth]{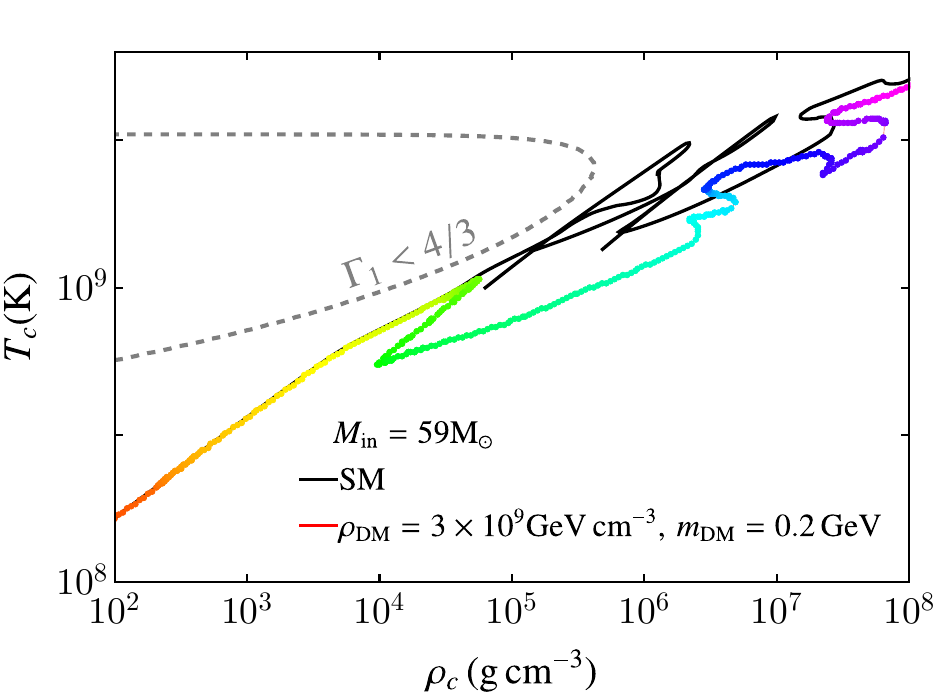}
    \includegraphics[width=.45\textwidth]{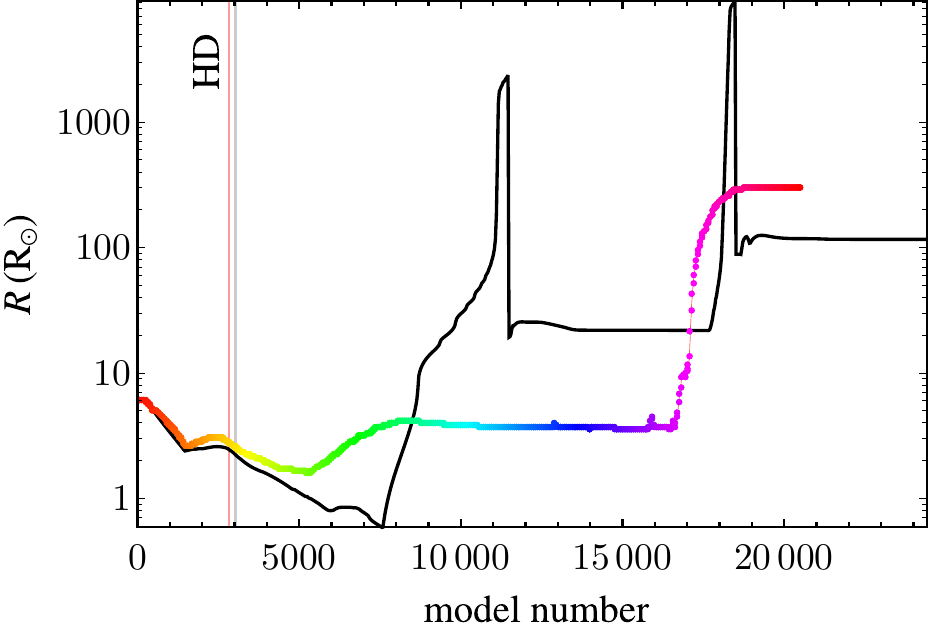}
    \caption{Example trajectory in core temperature and density of a $59\msun$ star in the SM and in the presence of DM energy injections, and the corresponding evolution of the radius with model number in the simulation.~The rainbow shading of the DM graph indicates the model number, for ease of comparison of the two panels.~As expected, the trajectory is unaltered before core helium depletion, indicated by ``HD'' (see text for details).~After HD, the injected energy can support the star against collapse, altering the trajectory in such a way that the star avoids the pulsational pair instability encountered in the SM.
    }
    \label{Fig:trajectories}
\end{figure}

The case where the DM mass is $\mdm=0.2$ GeV corresponds to a more diffuse DM profile where there is a significant amount of energy being injected in the star's outer layers.~As the temperature and density in these layers are too low for significant nuclear burning, DM heat can overpower SM heat at certain stages of the evolution.~After core helium depletion, this injection supports the star, preventing the contraction that would lead to C-burning in the SM.~The injected energy causes the star to expand, reducing the core temperature and density, moving the location of the track in the $\rho$--$T$ plane.~Examples are shown in figure~\ref{Fig:trajectories}.~The new track is akin to that of a lower mass star.~For sufficiently light objects, the pair-instability is avoided, and hence no PPISN/PISN occurs.~Nuclear burning eventually proceeds through the usual channels, and the star ultimately core collapses to a heavier black hole than predicted by the SM.~Heavier objects cannot evade the instability, and experience PPISN/PISN depending on the mass.~Therefore, a black hole mass gap still forms, but at higher masses than predicted by the SM.~

\section{Discussion}
\label{sec:conclusions}

In this work, we have investigated the effect of effect of dark matter annihilation on pair-instability supernovae.~We modeled the DM injection profile using the local thermal equilibrium assumption \eqref{eq:MBdist}.~We identified two separate phenomena depending on the DM mass.~If DM is heavier than $\sim\mathcal{O}(1)$ GeV, it is highly concentrated in the core.~The energy injection from its annihilation prolongs the lifetime of core helium burning, reducing the C/O ratio leading to more violent explosions that result in lower mass black holes being formed, and a reduction in the location of the lower edge of the upper black hole mass gap.~If DM is lighter than $\sim\mathcal{O}(1)$ GeV then it is diffused throughout the star and its energy injection from annihilation leads to a phase of partial DM support that resists the contraction that usually follows helium depletion.~Instead, the star expands, which reduces the core temperature and density.~Some stars that would have experienced the pair-instability evade it, leading to heavier black holes being formed and an increase in the location of the lower edge of the upper black hole mass gap.~We found that each of these effects --- more violent explosions for DM masses heavier than $\sim\mathcal{O}(1)$ GeV and evading the pair-instability for DM masses lower than $\sim\mathcal{O}(1)$ GeV --- are  exhibited when the ambient DM density $\rho_{\rm DM}\gtrsim10^9\, \rm GeV \, cm^{-3}$.~Such densities are only potentially realised in the very centers of DM spikes around galactic nuclei, and giant stars evolving in such backgrounds are expected to be exceedingly rare.~We therefore anticipate that the effects of DM will be difficult to observe in current GW catalogs.~{In practice, the low metallicity ($Z=10^{-5}$) stars we have simulated are unlikely to inhabit galactic centers, but we expect that DM will effect higher metallicity stars in a qualitatively similar manner.~The primary effect of metallicity is to alter the rate of mass loss to stellar winds during core helium burning ($\dot{M}\propto Z^{0.85}$).~Since DM injections do not alter stellar winds, we do not expect our conclusions to change for higher metallicity objects.  }


The effects of an additional non-nuclear energy injection on massive stars were studied recently by \cite{Ziegler:2020klg,Ziegler:2022apq}.~In those works, the authors injected a constant amount of energy per unit mass and time at each point in the star as a proxy for DM annihilation without specifying a profile.~The works conclude that the upper black hole mass gap is \textit{filled in} because stars can evade the instability.~In contrast, in this work we find a mass gap for all DM masses and ambient DM densities that we simulated.~We attempted to reproduce the results in these works based on the descriptions of the code  given in the papers, but all of our models failed to converge (unfortunately, the code used by \cite{Ziegler:2020klg,Ziegler:2022apq} is not publicly-available).~The scenario with a uniform heat injection throughout the star implies that the DM profile is diffuse, and this scenario is therefore most similar to our $\mdm=0.2$ GeV models.~Additionally, if one attempts to do this comparison, the uniform heat injection studied in
\cite{Ziegler:2020klg,Ziegler:2022apq} would correspond to ambient DM densities $\rho_{\rm DM}\gg 3\times10^9 \, \rm GeV \, cm^{-3}$, in which case more stars would evade the pair-instability but, following our results, one would still expect sufficiently heavy stars to traverse the instability region and experience PPISN/PISN, leading to a gap.~An exploration of a larger range of initial masses using their code could test this hypothesis.

As noted above, ambient DM densities $\rho_{\rm DM}>10^9  {\rm GeV \, cm^{-3}}$ could only be realised deep inside hypothetical DM spikes around SMBHs and, consequentially, one would expect the mass gap-filling objects formed in this way to be extremely rare.~In addition, the DM captured by a star is predicted to evaporate when $\mdm \lesssim 0.1$ GeV, so a near-uniform injection scenario would require a DM model with an additional attractive force between DM particles with Compton wavelength $\lambda_C\sim R_\odot$ in order to be viable.~Considering our results, we expect that mass-gap filling DM models  represent only a portion of the more general space of DM models, namely those with light DM masses, likely requiring an attractive dark force that prevents evaporation, and extremely high ambient DM densities.~

\dcedit{ Our study constitutes a first systematic exploration of the effect of DM heat injections with a LTE profile on the post-main sequence evolution of heavy stars.~DM may have an effect on cooler objects at lower ambient density.~We plan to explore this further in future work.\newline} 

\section*{Software}
MESA version~12778, MESASDK version 20200325, Asteria version 1, Mathematica version 12.

\section*{Acknowledgements}
We thank Rebecca Leane, Juri Smirnov, and Aaron Vincent for useful discussions.~DC is supported by the STFC under Grant No.~ST/T001011/1.~This material is based upon work supported by the National Science Foundation under Grant No.~2207880.~Our simulations were run on the University of Hawai\okina i's high-performance supercomputer MANA.~The technical support and advanced computing resources from University of Hawai\okina i Information Technology Services – Cyberinfrastructure, funded in part by the National Science Foundation MRI award \#1920304, are gratefully acknowledged.~

\bibliography{refs}

\end{document}